# Quantum teleportation between multiparties

Andrzej Grudka

Faculty of Physics, Adam Mickiewicz University

Umultowska 85, 61-614 Poznań, Poland

Abstract

The paper presents general protocols for quantum teleportation between multiparties. It is shown how $N$ parties can teleport $N$ unknown quantum states to $M$ other parties with the use of $N+M$ qudits in the maximally entangled state. It is also shown that a single pair of qudits in the maximally entangled state can be used to teleport two qudits in opposite directions simultaneously.

## I. Introduction

Quantum teleportation lies at the heart of quantum information. In the original scheme of Bennett *et al.* [1] two parties (Alice and Bob) share a pair of qubits in the maximally entangled state. In addition one party, e.g. Alice, has a qubit in an unknown quantum state. The two parties can perform a very simple protocol which enables a teleportation of the qubit in an unknown quantum state from Alice to Bob. In the same paper these results were generalized from qubits to qudits ($d$-dimensional quantum particles). On the other hand, Karlsson and Bourennane [2] and Hillery, Bužek and Barthiaume [3] proposed a different scheme where three parties (Alice, Bob 1, and Bob 2) share three qubits in the maximally entangled state. In their scheme one party (Alice) teleports an unknown quantum state of a qubit. At the end of the protocol the state of the original qubit is encoded in a two-qubit entangled state shared by Bob 1 and Bob 2 in such a way that none of them can reconstruct the original qubit by himself. However, when they cooperate and use only local quantum operations and classical communication, one of them can reconstruct the original qubit. This protocol can be easily generalized to qudits [4]. Quantum teleportation between multiparties with the use of qudits was also considered in [5, 6].

Let us now suppose that three parties (Bob 1, Bob 2, and Alice) share three qudits of dimensions $d$ in the maximally entangled state. In addition let two of them have qudits of dimensions $d_1$ and $d_2$ ($d_1 d_2 = d$) in unknown quantum states. We will present a protocol which enables them to teleport the states of their qudits to Alice with the use of the above mentioned qudits in the maximally entangled state. At the end of the protocol Alice will possess one qudit of dimension $d$, whose state encodes the states of the teleported qudits. The reverse problem will also be considered. Now let Alice has a qudit whose state encodes the states of two qudits of lower dimensions. It will be shown how Alice can teleport the first encoded qudit to Bob 1 and the second one to Bob 2. Both protocols will be generalized to the case of more than three parties. Finally, we will consider quantum teleportation between many senders and many receivers of quantum information. There will also be shown that a single pair of entangled qudits can serve as a two-way quantum channel. Namely, if Alice and Bob share such a pair then Alice can teleport a qudit of dimension $d_1$ to Bob and Bob can teleport a qudit of dimension $d_2$ to Alice with the use of one pair of qudits of dimension $d$ in the maximally entangled state.

The paper is organized as follows. Part II describes quantum teleportation from one party to many parties while part III deals with quantum teleportation from many parties to one party. Finally in part IV quantum teleportation from many senders to many receivers is described. For pedagogical reasons simple examples are presented first and then the results are generalized.

## II. Quantum teleportation: many to one

Let three parties (Bob 1, Bob 2, and Alice) share three qudits in the maximally entangled state of the form

$$|\Psi\rangle = \frac{1}{\sqrt{d}} \sum_{k=0}^{d-1} |k\rangle_1 |k\rangle_2 |k\rangle_3, \quad (1)$$

where $d$ is a dimension of each qudit. We will assume that $d$ can be written as a product of two integers $d_1$ and $d_2$ ($d = d_1 d_2$) where both $d_1$ and $d_2$ are greater than 1. Let in addition Bob 1 and Bob 2 have qudits of dimension $d_1$ and $d_2$, respectively, in unknown quantum states. Because both $d_1$ and $d_2$ are smaller than $d$ each of these qudits can be encoded in the qudit of dimension $d$. We are thus free to assume that Bob 1 has a qudit of dimension $d$ in the state

$$|\alpha^1\rangle = \sum_{k_1=0}^{d_1-1} \alpha^1_{k_1} |k_1\rangle, \quad (2)$$

while Bob 2 has a qudit of dimension $d$ in a state

$$|\alpha^2\rangle = \sum_{k_2=0}^{d_2-1} \alpha^2_{k_2} |k_2 d_1\rangle. \quad (3)$$

Below we present the protocol which enables Bob 1 and Bob 2 to teleport their qudits "$\alpha^1$" and "$\alpha^2$" to Alice with the use of three qudits in the state "$\Psi$".

Bob 1 performs the operations given below.

(B1.2) He makes the unitary operation on his qudit "$\alpha^1$"

$$|k_1\rangle \to \frac{1}{\sqrt{d_2}} \sum_{k_2=0}^{d_2-1} |k_2 d_1 + k_1\rangle. \quad (4)$$

The new state of his qudit is

$$|\alpha^1\rangle^{(1)} = \frac{1}{\sqrt{d_2}} \sum_{k_2=0}^{d_2-1}\sum_{k_1=0}^{d_1-1} \alpha_{k_1}^1 |k_2 d_1 + k_1\rangle. \qquad (5)$$

(B1.3) He measures the qudits "$\alpha^1$" and "1" from "$\Psi$" in the generalized Bell basis

$$|\psi_{mn}\rangle = \frac{1}{\sqrt{d}} \sum_{l=0}^{d-1} \omega^{nl} |l\rangle |l \oplus m\rangle, \qquad (6)$$

where $\omega = \exp(i2\pi/d)$ and $\oplus$ denotes addition modulo $d$. If the result of his measurement is $\psi_{mn}$ then Bob 2 and Alice share two qudits in the state

$$|\Psi^{23}\rangle^{(1)} = \frac{1}{\sqrt{d_2}} \sum_{k_2=0}^{d_2-1}\sum_{k_1=0}^{d_1-1} \alpha_{k_1}^1 \omega^{n(k_2 d_1 + k_1)} |(k_2 d_1 + k_1) \oplus m\rangle_2 |(k_2 d_1 + k_1) \oplus m\rangle_3. \qquad (7)$$

(B1.4) He sends the result of his measurement to Alice and Bob 2.

(A.1) Alice performs the unitary operation

$$\omega^{n(k_2 d_1 + k_1)} |(k_2 d_1 + k_1) \oplus m\rangle_3 \to |k_2 d_1 + k_1\rangle_3. \qquad (8)$$

(B2.1) Bob 2 performs the unitary operation

$$|(k_2 d_1 + k_1) \oplus m\rangle_2 \to |k_2 d_1 + k_1\rangle_2. \qquad (9)$$

Now Bob 2 and Alice share two qudits in the state

$$|\Psi^{23}\rangle^{(2)} = \frac{1}{\sqrt{d_2}} \sum_{k_2=0}^{d_2-1}\sum_{k_1=0}^{d_1-1} \alpha_{k_1}^1 |k_2 d_1 + k_1\rangle_2 |k_2 d_1 + k_1\rangle_3. \qquad (10)$$

Bob 2 performs the operations given below.

(B2.2) He makes the unitary operation on his qudit "$\alpha^2$"

$$|k_2 d_1\rangle \to \frac{1}{\sqrt{d_1}} \sum_{k_1=0}^{d_1-1} |k_2 d_1 + k_1\rangle. \qquad (11)$$

The new state of his qudit is

$$|\alpha^2\rangle^{(1)} = \frac{1}{\sqrt{d_1}} \sum_{k_2=0}^{d_2-1}\sum_{k_1=0}^{d_1-1} \alpha_{k_2}^2 |k_2 d_1 + k_1\rangle. \qquad (12)$$

(B2.3) He measures the qudits "$\alpha^2$" and "2" from "$\Psi$" in the generalized Bell basis.

If the result of his measurement is $\psi_{mn}$ then Alice has the qudit in the state

$$\left|\Psi^3\right\rangle^{(1)} = \sum_{k_2=0}^{d_2-1}\sum_{k_1=0}^{d_1-1} \alpha^1_{k_1}\alpha^2_{k_2}\omega^{n(k_2 d_1+k_1)}\left|(k_2 d_1+k_1)\oplus m\right\rangle_3. \tag{13}$$

(B2.4) He sends the result of his measurement to Alice.

(A.2) Alice performs the unitary operation (8)

Now Alice has the qudit in the state

$$\left|\Psi^3\right\rangle^{(2)} = \sum_{k_2=0}^{d_2-1}\sum_{k_1=0}^{d_1-1} \alpha^1_{k_1}\alpha^2_{k_2}\left|k_2 d_1+k_1\right\rangle_3. \tag{14}$$

This state contains the same information as the states of qudits "$\alpha^1$" and "$\alpha^2$", so the teleportation has been performed successfully.

These results can be generalized to the case of more than three parties (Bob 1, Bob 2, …, Bob $N$ and Alice). ($N+1$) parties share ($N+1$) qudits in the maximally entangled state of the form

$$\left|\Psi\right\rangle = \frac{1}{\sqrt{d}}\sum_{k=0}^{d-1}\left|k\right\rangle_1\left|k\right\rangle_2\ldots\left|k\right\rangle_N\left|k\right\rangle_{N+1}. \tag{15}$$

Bob $i$ has a qudit of dimension $d_i$ encoded in the qudit of dimension $d$ ($d = d_1 d_2 \ldots d_N$) in the following way

$$\left|\alpha^i\right\rangle = \sum_{k_i=0}^{d_i-1}\alpha^i_{k_i}\left|k_i p_i\right\rangle, \tag{16}$$

where $p_1 = 1$ and $p_i = d_1 \ldots d_{i-1}$ ($i \neq 1$).

Below we present the protocol which enables all Bobs to teleport their qudits to Alice.

Bob $i$ has to perform the following operations.

(B$i$.1) He makes the unitary operation

$$\left|(k_N p_N + \ldots + k_2 p_2 + k_1)\oplus m\right\rangle_i \to \left|k_N p_N + \ldots + k_2 p_2 + k_1\right\rangle_i \tag{17}$$

for all results of measurements he receives from Bob 1, Bob 2,..., Bob $i-1$. Bob 1 does not perform this step.

(B$i$.2) He makes the unitary operation

$$|k_i p_i\rangle_i \rightarrow \sqrt{\frac{d_i}{d}} \sum_{m_1}^{d_1-1} \ldots \sum_{m_{i-1}}^{d_{i-1}-1} \sum_{m_{i+1}}^{d_{i+1}-1} \ldots \sum_{m_N}^{d_N-1} |m_N p_N + \ldots m_{i+1} p_{i+1} + k_i p_i + m_{i-1} p_{i-1} + \ldots m_1\rangle_i. \quad (18)$$

(B$i$.3) He measures the qudits "$\alpha^2$" and "$i$" from "$\Psi$" in the generalized Bell basis

(B$i$.4) He sends the result of his measurement to Alice and Bob $i+1$, Bob $i+2$,..., Bob $N$.

Alice has to perform the unitary operation

$$\omega^{n(k_N p_N + \ldots + k_2 p_2 + k_1)} |(k_N p_N + \ldots + k_2 p_2 + k_1) \oplus m\rangle_{N+1} \rightarrow |k_N p_N + \ldots + k_2 p_2 + k_1\rangle_{N+1}. \quad (19)$$

for all results of measurements she receives from Bob 1, Bob 2,...,Bob $N$.

At the end of this protocol Alice has the qudit in the state

$$|\Psi^{N+1}\rangle = \sum_{k_1=0}^{d_1-1} \sum_{k_2=0}^{d_2-1} \ldots \sum_{k_N=0}^{d_N-1} \alpha^1_{k_1} \alpha^2_{k_2} \ldots \alpha^N_{k_N} |k_N p_N + \ldots + k_2 p_2 + k_1 p_1\rangle_{N+1}. \quad (20)$$

## III. Quantum teleportation: one to many

Now we will consider the reverse problem. Let three parties (Alice 1, Alice 2, and Bob) share three qudits in the maximally entangled state (1). Let Bob have two qudits "$\alpha^1$" and "$\alpha^2$" of dimensions $d_1$ and $d_2$ encoded in one qudit of dimension $d$ ($d = d_1 d_2$) in the following way

$$|\alpha^{1,2}\rangle = \sum_{k_2=0}^{d_2-1} \sum_{k_1=0}^{d_1-1} \alpha^1_{k_1} \alpha^2_{k_2} |k_2 d_1 + k_1\rangle. \quad (21)$$

Below we present the protocol which enables Bob to teleport his qudits "$\alpha^1$" and "$\alpha^2$" to Alice 1 and Alice 2 respectively with the use of three qudits in the state "$\Psi$".

Bob performs the operations given below

(B.1) He measures the qudits "$\alpha^{1,2}$" and "3" from "$\Psi$" in the generalized Bell basis. If the result of his measurement is $\psi_{mn}$ then Alice 1 and Alice 2 share two qudits in the state

$$\left|\Psi^{1,2}\right\rangle^{(1)} = \sum_{k_2=0}^{d_2-1}\sum_{k_1=0}^{d_1-1}\alpha_{k_1}^1\alpha_{k_2}^2\omega^{n(k_2d_1+k_1)}\left|(k_2d_1+k_1)\oplus m\right\rangle_1\left|(k_2d_1+k_1)\oplus m\right\rangle_2. \tag{22}$$

(B.2) He sends the result of his measurement to Alice 1 and Alice 2.

(A1.1) Alice 1 performs the unitary operation

$$\omega^{n(k_2d_1+k_1)}\left|(k_2d_1+k_1)\oplus m\right\rangle_1 \rightarrow \left|k_2d_1+k_1\right\rangle_1. \tag{23}$$

(A2.1) Alice 2 performs the unitary operation

$$\left|(k_2d_1+k_1)\oplus m\right\rangle_2 \rightarrow \left|k_2d_1+k_1\right\rangle_2. \tag{24}$$

Now Alice 1 and Alice 2 share two qudits in the state

$$\left|\Psi^{1,2}\right\rangle^{(2)} = \sum_{k_2=0}^{d_2-1}\sum_{k_1=0}^{d_1-1}\alpha_{k_1}^1\alpha_{k_2}^2\left|k_2d_1+k_1\right\rangle_1\left|k_2d_1+k_1\right\rangle_2. \tag{25}$$

Alice 1 performs the operations given below

(A1.2) She makes the unitary operation

$$\left|k_2d_1+k_1\right\rangle_1 \rightarrow \frac{1}{\sqrt{d_2}}\sum_{m=0}^{d_2-1}\omega^{k_2d_1m}\left|md_1+k_1\right\rangle_1. \tag{26}$$

The new state is

$$\left|\Psi^{1,2}\right\rangle^{(3)} = \frac{1}{\sqrt{d_2}}\sum_{k_2=0}^{d_2-1}\sum_{k_1=0}^{d_1-1}\sum_{m=0}^{d_2-1}\omega^{k_2d_1m}\alpha_{k_1}^1\alpha_{k_2}^2\left|md_1+k_1\right\rangle_1\left|k_2d_1+k_1\right\rangle_2. \tag{27}$$

(A1.3) She measures the projectors

$$P_m = \sum_{k_1=0}^{d_1-1}\left|md_1+k_1\right\rangle_{11}\left\langle md_1+k_1\right|. \tag{28}$$

If the result of her measurement is $P_m$ then Alice 1 and Alice 2 share two qudits in the state

$$\left|\Psi^{1,2}\right\rangle^{(4)} = \sum_{k_2=0}^{d_2-1}\sum_{k_1=0}^{d_1-1}\omega^{k_2d_1m}\alpha_{k_1}^1\alpha_{k_2}^2\left|md_1+k_1\right\rangle_1\left|k_2d_1+k_1\right\rangle_2. \tag{29}$$

(A1.4) She sends the result of her measurement to Alice 2.

(A1.5) She makes the unitary operation

$$|md_1 + k_1\rangle_1 \to |k_1\rangle_1. \tag{30}$$

Alice 2 performs the operations given below.

(A2.6) She makes the unitary operation (This step can be performed after step (A2.5), however for clarity I have put it here.)

$$\omega^{k_2 d_1 m}|k_2 d_1 + k_1\rangle_2 \to |k_2 d_1 + k_1\rangle_2. \tag{31}$$

The new state is

$$|\Psi^{1,2}\rangle^{(5)} = \sum_{k_2=0}^{d_2-1}\sum_{k_1=0}^{d_1-1} \alpha^1_{k_1}\alpha^2_{k_2}|k_1\rangle_1|k_2 d_1 + k_1\rangle_2. \tag{32}$$

(A2.2) She makes the unitary operation

$$|k_2 d_1 + k_1\rangle_2 \to \frac{1}{\sqrt{d_1}}\sum_{m=0}^{d_1-1}\omega^{k_1 d_2 m}|k_2 d_1 + m\rangle_2. \tag{33}$$

The new state is

$$|\Psi^{1,2}\rangle^{(6)} = \frac{1}{\sqrt{d_1}}\sum_{k_2=0}^{d_2-1}\sum_{k_1=0}^{d_1-1}\sum_{m=0}^{d_2-1}\omega^{k_1 d_2 m}\alpha^1_{k_1}\alpha^2_{k_2}|k_1\rangle_1|k_2 d_1 + m\rangle_2. \tag{34}$$

(A2.3) She measures the projectors

$$P_m = \sum_{k_2=0}^{d_2-1}|k_2 d_1 + m\rangle_{2\,2}\langle k_2 d_1 + m|. \tag{35}$$

If the result of his measurement is $P_m$ then Alice 1 and Alice 2 share two qudits in the state

$$|\Psi^{1,2}\rangle^{(7)} = \sum_{k_2=0}^{d_2-1}\sum_{k_1=0}^{d_1-1}\omega^{k_1 d_2 m}\alpha^1_{k_1}\alpha^2_{k_2}|k_1\rangle_1|k_2 d_1 + m\rangle_2. \tag{36}$$

(A2.4) She sends the result of her measurement to Alice 2.

(A2.5) She performs the unitary operation

$$|k_2 d_1 + m\rangle_1 \to |k_2 d_1\rangle_1. \tag{37}$$

(A1.6) Alice 1 performs the unitary operation

$$\omega^{k_1 d_2 m}|k_1\rangle_2 \to |k_1\rangle_2. \qquad (38)$$

Now Alice 1 and Alice 2 have two qudits in the state

$$|\Psi^{1,2}\rangle^{(8)} = \sum_{k_1=0}^{d_1-1}\alpha^1_{k_1}|k_1\rangle_1 \sum_{k_2=0}^{d_2-1}\alpha^2_{k_2}|k_2 d_1\rangle_2. \qquad (39)$$

It should be noted that the corresponding steps for Alice 1 and Alice 2 (e.g. A1.3 and A2.3) can be performed in parallel.

These results can also be generalized to the case of $(N+1)$ parties (, and Alice 1, Alice 2, …, Alice $N$, and Bob). $(N+1)$ parties share $(N+1)$ qudits in the maximally entangled state (15). Bob has $N$ qudits "$\alpha^1$", "$\alpha^2$",…, "$\alpha^N$" of dimensions $d_1$, $d_2$, …, $d_N$ encoded in one qudit of dimension $d$ in the following way

$$|\alpha^{1,2,\ldots,N}\rangle = \sum_{k_1=0}^{d_1-1}\sum_{k_2=0}^{d_2-1}\cdots\sum_{k_N=0}^{d_N-1}\alpha^1_{k_1}\alpha^2_{k_2}\ldots\alpha^N_{k_N}|k_N p_N + \ldots + k_2 p_2 + k_1 p_1\rangle. \qquad (40)$$

Below we present the protocol which enables Bob to teleport $N$ qudits to $N$ remaining parties (one qudit to one party). Because the generalization of steps performed by Bob and the first step performed by each Alice is straightforward we consider only the generalization of all the next steps performed by each Alice.

(A$i$.2) Alice $i$ makes the unitary operations

$$|k_N p_N + \ldots k_{i+1}p_{i+1} + k_i p_i + k_{i-1}p_{i-1} + \ldots k_1\rangle_i \to$$
$$\sum_{m_1}^{d_1-1}\ldots\sum_{m_{i-1}}^{d_{i-1}-1}\sum_{m_{i+1}}^{d_{i+1}-1}\ldots\sum_{m_N}^{d_N-1}\omega_N^{m_N^i}\ldots\omega_{i+1}^{m_{i+1}^i}\omega_{i-1}^{m_{i-1}^i}\ldots\omega_1^{m_1^i}|m_N^i p_N + \ldots m_{i+1}^i p_{i+1} + k_i p_i + m_{i-1}^i p_{i-1} + \ldots m_1^i\rangle_i, \qquad (41)$$

where $\omega_i = \omega^{d/d_i}$.

(A$i$.3) Alice $i$ measures the projectors

$$P^i_{m_N^i \ldots m_{i+1}^i m_{i-1}^i \ldots m_1^i} =$$
$$\sum_{k_i=0}^{d_i-1}|m_N^i p_N + \ldots m_{i+1}^i p_{i+1} + k_i p_i + m_{i-1}^i p_{i-1} + \ldots m_1^i\rangle_{i\,i}\langle m_N^i p_N + \ldots m_{i+1}^i p_{i+1} + k_i p_i + m_{i-1}^i p_{i-1} + \ldots m_1^i|.$$

(42)

(A$i$.4) Alice $i$ sends the result $m_j^i$ of her measurement to Alice $j$.

(A$i$.5) Alice $i$ makes the unitary operation

$$\left| m_N^i p_N + ... m_{i+1}^i p_{i+1} + k_i p_i + m_{i-1}^i p_{i-1} + ... m_1^i \right\rangle_i \to \left| k_i p_i \right\rangle_i. \qquad (43)$$

(A$i$.6) Alice $i$ performs the unitary operation

$$\omega_i^{m_i^j} \left| k_N p_N + ... k_{i+1} p_{i+1} + k_i p_i + k_{i-1} p_{i-1} + ... k_1 \right\rangle_i \to$$
$$\left| k_N p_N + ... k_{i+1} p_{i+1} + k_i p_i + k_{i-1} p_{i-1} + ... k_1 \right\rangle_i \qquad (44)$$

for all results of measurements $j \neq i$ she receives from Alice 1,..., Alice $j-1$, Alice $j+1$,...,Alice $N$.

At the end of this protocol Alice 1, Alice 2,...,Alice $N$ have $N$ qudits in the state

$$\left| \Psi^{1,2,...,N} \right\rangle = \sum_{k_1=0}^{d_1-1} \alpha_{k_1}^1 \left| k_1 p_1 \right\rangle_1 \sum_{k_2=0}^{d_2-1} \alpha_{k_2}^2 \left| k_2 p_2 \right\rangle_2 ... \sum_{k_N=0}^{d_N-1} \alpha_{k_N}^N \left| k_N p_N \right\rangle_N. \qquad (45)$$

### IV. Quantum teleportation: many to many

Let us now assume that $2N$ parties ($N$ Bobs, who are senders of quantum information and $N$ Alices, who are receivers) share $2N$ qudits of dimension $d$ in the maximally entangled state. Let $N$ of them (each Bob) have qudits of dimension $d_1$, $d_2$, ..., $d_N$ and $N+1$ parties (each Bob and Alice 1) perform the steps described in part II. In addition let each Bob send the result of his measurement in step (B$i$.2) not only to Alice 1 but also to Alice 2, Alice 3,..., Alice $N$. If Alice 2, Alice 3,..., Alice $N$ perform the unitary operation given by Eq. (17) on their qudits for all results of measurements they receive from Bobs then $N$ Alices will share $N$ qudits in the state

$$\left| \alpha^{1,2,...,N} \right\rangle = \sum_{k_1=0}^{d_1-1} \sum_{k_2=0}^{d_2-1} ... \sum_{k_N=0}^{d_N-1} \alpha_{k_1}^1 \alpha_{k_2}^2 ... \alpha_{k_N}^N \times$$
$$\left| k_N p_N + ... + k_2 p_2 + k_1 p_1 \right\rangle \left| k_N p_N + ... + k_2 p_2 + k_1 p_1 \right\rangle ... \left| k_N p_N + ... + k_2 p_2 + k_1 p_1 \right\rangle \qquad (46)$$

They can now perform the steps described in part III. At the end of the protocol these $N$ Alices will have the state (45). Thus the quantum states have been teleported from Bobs to Alices. Because some of the parties can be in fact the same party (e.g. Alice 1 = Bob 2) we have a kind of quantum teleportation network which enables exchange of quantum information between $M \leq 2N$ parties as long as all the quantum information can be encoded in a qudit of dimension $d$. It should be noted that in the case when we have $M$ (not $2N$) parties the parties should share only $M$ qudits in the maximally entangled state, because one can entangle additional qudits with the use of $XOR$ gate defined as

$$XOR|k\rangle|l\rangle = |k\rangle|l \oplus k\rangle. \tag{47}$$

Let us now discuss very simple example. We have two parties: Bob 1 = Alice 2 (I will call him/her Bob) and Bob 2 = Alice 1 (I will call him/her Alice). They share two qudits of dimension $d$ in the maximally entangled state. Bob has a qudit of dimension $d_1$ and wants to teleport it to Alice. Alice has a qudit of dimension $d_2$ ($d_1 d_2 = d$) and wants to teleport it to Bob. They will achieve this aim if both Bob and Alice entangle additional qudits of dimension $d$ to their two qudits in the maximally entangled state and then perform the protocol just described. One thus sees that a single pair of maximally entangled qudits can serve as a two-way quantum channel in the sense that the same pair can be used to teleport one qudit from Alice to Bob and one qudit from Bob to Alice.

## V. Summary

A general protocol for quantum teleportation between multiple parties with the use of qudits has been proposed. The protocol enables exchange of quantum information between $M$ parties with the use of the maximally entangled state of $M$ qudits. It enables exchange quantum information even if it is not possessed by one party. Moreover, different parts of quantum information can be transmitted to different parties.

## Acknowledgements

I would like to thank the State Committee for Scientific Research for financial support under Grant No. 0 T00A 003 23. I would also like to thank the European Science Foundation for a Short Scientific Visit under QIT Programme. I would like to thank Artur Ekert and Antoni Wójcik for discussions.